# Expansions of the solutions of the biconfluent Heun equation in terms of incomplete Beta and Gamma functions

T.A. Ishkhanyan[1,2], Y. Pashayan-Leroy[3], M.R. Gevorgyan[1,3], C. Leroy[3], and A.M. Ishkhanyan[1]

[1]Institute for Physical Research of NAS of Armenia, Ashtarak, 0203 Armenia
[2]Moscow Institute of Physics and Technology, Dolgoprudny, 141700 Russia
[3]Laboratoire Interdisciplinaire Carnot de Bourgogne, CNRS UMR 6303, Université de Bourgogne, BP 47870, 21078 Dijon, France

Starting from equations obeyed by functions involving the first or the second derivatives of the biconfluent Heun function, we construct two expansions of the solutions of the biconfluent Heun equation in terms of incomplete Beta functions. The first series applies single Beta functions as expansion functions, while the second one involves a combination of two Beta functions. The coefficients of expansions obey four- and five-term recurrence relations, respectively. It is shown that the proposed technique is potent to produce series solutions in terms of other special functions. Two examples of such expansions in terms of the incomplete Gamma functions are presented.

## 1. Introduction

The biconfluent Heun equation [1-2] is widely involved in different domains of contemporary pure and applied sciences such as quantum mechanics, general relativity, solid state physics, atomic, molecular and optical physics, chemistry, etc. (see, e.g. [3-9]). A recent example is the inverse square root potential [10], a member of the biconfluent Heun potentials [11], which describes a less singular interaction than the Coulomb potential.

Though the properties of the biconfluent Heun equation have been studied by many authors (see, e.g., [12-23]), however, there are many open problems in the theory of this equation. For instance, for a long time the applications have been restricted to the polynomial solutions only. An extension to the non-polynomial ones is an interesting and important challenge [9]. Indeed, as we have recently shown, by expanding the solutions in terms of more advanced mathematical functions rather than in terms of powers may result in tractable algorithms for deriving new analytic forms of square-integrable non-polynomial solutions of the Schrödinger equation [10]. In the present paper we make a step in this direction by constructing two expansions of the solutions of the biconfluent Heun equation in terms of the incomplete Beta functions and two expansions in terms of the incomplete Gamma functions.

The approach we apply to construct these expansions is as follows. We consider, following the lines of [24-31], an expansion of an auxiliary function involving a *derivative* of the biconfluent Heun function in terms of some elementary or special functions. A notable feature of these equations is that in general they have at least one more regular singularity, as compared with the biconfluent Heun equation. The position of this extra singularity is defined by the accessory parameter of the biconfluent Heun equation (i.e., the parameter that originates from the accessory parameter of the ancestor general Heun equation [32]), and in general it may be located at any point of the extended complex $z$-plane. Having the expansion of the mentioned auxiliary function, by further integration(s), we derive an expansion of the biconfluent Heun function in terms of resulting special functions or combinations of such functions.

Below we discuss two different types of expansions in terms of incomplete Beta functions, which are derived using a Frobenius and a Beta function expansions for auxiliary functions that involve the first and the second derivatives of the biconfluent Heun function, respectively. The coefficients of expansions obey four- and five-term recurrence relations, respectively. Finally, we present two expansions in terms of the incomplete Gamma function. The constructed expansions apply to arbitrary set of the involved parameters of the biconfluent Heun equation with proviso $\alpha \neq 0$ and $q \neq 0$.

According to the general theory [1-2, 12], the biconfluent Heun equation has four irreducible parameters. A canonical form of this equation adopted in [1] is written as

$$\frac{d^2u}{dz^2} + \frac{1+\alpha-\beta z-2z^2}{z}\frac{du}{dz} + \frac{(\gamma-\alpha-2)z-(\delta+(1+\alpha)\beta)/2}{z}u = 0. \qquad (1)$$

As already mentioned above, the singularities of this equation are located at $z=0$ (regular singularity) and at $z=\infty$ (irregular singularity of rank 2). A different form also involving four independent parameters is adopted in [2]:

$$\frac{d^2u}{dz^2} - \left(\frac{\gamma}{z}+\delta+z\right)\frac{du}{dz} + \frac{\alpha z - q}{z}u = 0. \qquad (2)$$

Depending on the particular developments of interest and corresponding theoretical background, other canonical forms may be suitable, as stated by the authors of [1]. For the sake of generality, in the present paper we adopt the following form of this equation:

$$\frac{d^2u}{dz^2} + \left(\frac{\gamma}{z}+\delta+\varepsilon z\right)\frac{du}{dz} + \frac{\alpha z - q}{z}u = 0. \qquad (3)$$

It is readily seen that the above two forms as well as other forms applied in the literature are derived from this form by straightforward specifications of the involved parameters.

A few remarks concerning some elementary cases of the biconfluent Heun equation are relevant. First of all, we note that Eq. (3) is immediately reduced to the Kummer confluent hypergeometric equation if $\varepsilon = 0$ and $\alpha = 0$. Furthermore, in fact, the case $\varepsilon = 0$ is always reducible, irrespective of the value of $\alpha$, because in this case Eq. (3) is reduced to the confluent hypergeometric equation by a simple transformation of the dependent variable $u = e^{sz} v(z)$. The general solution of the equation is then written as

$$u = e^{sz} \left[ C_1 \cdot {}_1F_1\left((q-\gamma s)/s_0; \gamma; s_0 z\right) + C_2 U\left((q-\gamma s)/s_0; \gamma; s_0 z\right) \right], \qquad (4)$$

where ${}_1F_1$ and $U$ are the Kummer and the Tricomi confluent hypergeometric functions, respectively, $C_{1,2}$ are constants and

$$s = -(\delta + s_0)/2, \quad s_0 = \pm\sqrt{\delta^2 - 4\alpha}. \qquad (5)$$

Another known case when the solution of Eq. (3) is written in terms of the confluent hypergeometric functions (this time, of the argument $-\varepsilon z^2/2$) is the case $\delta = q = 0$ [1]. Finally, a simple case, in a sense degenerate, is the case $\alpha = 0$ and $q = 0$, when the general solution of the biconfluent Heun equation is readily written in quadratures:

$$u = C_1 + C_2 \int e^{-\delta z - \varepsilon z^2/2} z^{-\gamma} dz, \quad C_{1,2} = \text{const}. \qquad (6)$$

Taking into account above observations, below we suppose that $\varepsilon \neq 0$, as well as $\alpha$ and $q$ are not simultaneously zero.

## 2. Expansions in terms of the incomplete Beta functions

To demonstrate the technique we apply for construction of series solutions in terms of incomplete Beta functions, consider the following representation of the first derivative of a solution of Eq. (3):

$$\frac{du}{dz} = z^{-\gamma} v(z). \qquad (7)$$

If now the function $v(z)$ allows an expansion of the form

$$v(z) = \sum_{n=0}^{+\infty} c_n (z-s)^{\mu+n}, \qquad (8)$$

then the term-by-term integration of Eq. (7) produces the following expansion in terms of incomplete Beta functions ($|z| \leq |z_0|$):

$$u = C_0 + \sum_{n=0}^{+\infty} c_n (-s)^n B\left(1-\gamma,\ 1+n+\mu;\ \frac{z}{s}\right). \qquad (9)$$

A more elaborate example, this time involving incomplete Beta functions already in Eq. (8), is constructed using a possible expansion for the function $v(z)$ of the form:

$$v(z) = C_1 + \sum_{n=0}^{+\infty} c_n B(a_n, b_n; z/s). \tag{10}$$

Then, the term-by-term integration produces an expansion in terms of combinations of incomplete Beta functions:

$$u(z) = C_0 + C_1 \frac{z^{1-\gamma}}{1-\gamma} + \sum_{n=0}^{+\infty} \frac{c_n}{1-\gamma} \left( z^{1-\gamma} B(a_n, b_n; z/s) - s^{1-\gamma} B(a_n + 1 - \gamma, b_n; z/s) \right). \tag{11}$$

Note that the integration constant $C_0$ and $C_1$ in above expansions are not arbitrary; rather, they should be appropriately chosen in order to achieve a consistent solution.

In the force of the presented approach, it is understood that the task now is to look at different equations obeyed by functions involving the derivatives of the biconfluent Heun function and to construct expansions for the latter functions starting from these equations.

To be more specific, consider, for instance, the details of derivation of the expansion (9). It is readily shown, e.g., by dividing Eq. (3) by $(\alpha z - q)/z$ and further differentiating, that a differential equation obeyed by the function $v(z) = z^\gamma du/dz$ is written as:

$$\frac{d^2 v}{dz^2} + \left( \frac{1-\gamma}{z} + \delta + \varepsilon z - \frac{\alpha}{\alpha z - q} \right) \frac{dv}{dz} + \frac{\Pi(z)}{z(\alpha z - q)} v = 0, \tag{12}$$

where $\Pi(z)$ is a quadratic polynomial in $z$:

$$\Pi(z) = \alpha(\alpha + \varepsilon - \gamma\varepsilon)z^2 - (\alpha(2q + \gamma\delta) + q\varepsilon(2-\gamma))z + q(q + (\gamma-1)\delta). \tag{13}$$

As compared with Eq. (3), it is immediately seen that this equation has an additional regular singularity at $z_0 = q/\alpha$. Let now $\alpha \neq 0$ and $q \neq 0$ so that the additional singular point is a finite point of the complex $z$-plane, not located at the origin: $z_0 \neq 0$. Then, taking the Frobenius solution of Eq. (12) in the neighborhood of this singularity:

$$v = (z - z_0)^\mu \sum_{n=0}^{+\infty} a_n^{(z_0)} (z - z_0)^n, \tag{14}$$

we get the expansion (8), and further term-by-term integrating Eq. (7) we arrive at the expansion (9) finally written as ($|z| \leq |z_0|$)

$$u = C_0 + \sum_{n=0}^{+\infty} a_n^{(z_0)} (-z_0)^n B\left( 1-\gamma,\, 1+n+\mu;\, \frac{z}{z_0} \right). \tag{15}$$

Substituting this expansion into Eq. (3) and taking the limit $z \to 0$ we readily find that $C_0 = 0$ if $\mathrm{Re}(1-\gamma) > 0$. The successive coefficients of the constructed expansion obey a four-term recurrence relation:

$$S_n a_n^{(z_0)} + R_{n-1} a_{n-1}^{(z_0)} + Q_{n-2} a_{n-2}^{(z_0)} + P_{n-3} a_{n-3}^{(z_0)} = 0, \qquad (16)$$

where

$$S_n = z_0(n+\mu)(n+\mu-2), \qquad (17)$$

$$R_n = z_0(\delta + z_0 \varepsilon)(n+\mu-1) + (n+\mu)(n+\mu-1-\gamma), \qquad (18)$$

$$Q_n = -\gamma(\delta + z_0 \varepsilon) + (\delta + 2z_0 \varepsilon)(n+\mu), \qquad (19)$$

$$P_n = \alpha + \varepsilon(n+\mu+1-\gamma). \qquad (20)$$

The series is left-hand side terminated at $n=0$ if $S_0 = 0$, i.e., if $\mu = 0$ or $\mu = 2$. These exponents differ by an integer, and it is readily checked that only the greater exponent $\mu = 2$ produces a consistent power-series expansion; the second independent solution requires a logarithmic term. The series terminates from the right-hand side if three successive coefficients vanish for some $N = 1, 2, \ldots$: $a_N \neq 0$, $a_{N+1} = a_{N+2} = a_{N+3} = 0$. From the condition $a_{N+3} = 0$ we find that the termination is possible if $P_N = 0$, i.e., if

$$\alpha = -\varepsilon(N+\mu+1-\gamma), \quad \mu = 2. \qquad (21)$$

The remaining two equations, $a_{N+1} = 0$ and $a_{N+2} = 0$, then impose two more restrictions on the parameters of the bi-confluent Heun equation.

Consider now how to derive an expansion of the form of Eq. (11), in terms of combinations of incomplete Beta functions. It is readily verified that the first derivative $du(z)/dz = v(z)$ of the biconfluent Heun function obeys the differential equation

$$v_{zz} + \left(\frac{1+\gamma}{z} + \delta + \varepsilon z - \frac{\alpha}{\alpha z - q}\right) v_z + \frac{(\alpha+\varepsilon)z(\alpha z - 2q) + (q^2 - \delta q - \alpha \gamma)}{z(\alpha z - q)} v = 0, \qquad (22)$$

which has an additional regular singularity at $z_0 = q/\alpha$. We now need to construct a Beta function expansion of a solution of this equation. As we have seen above, such an expansion can be constructed by passing to the equation obeyed by the function $w(z) = z^{1+\gamma} dv(z)/dz$. For $(\alpha+\varepsilon)\alpha \neq 0$ this equation is written as

$$w_{zz} + \left(-\frac{\gamma}{z} + \delta + \varepsilon z - \frac{1}{z-z_1} - \frac{1}{z-z_2}\right) w_z + \frac{P_3(z)}{z(z-z_1)(z-z_2)} w = 0, \qquad (23)$$

where $P_3(z)$ is a cubic polynomial and $z_{1,2}$ are the roots of the quadratic equation

$$(\alpha+\varepsilon)z(\alpha z - 2q) + (q^2 - \delta q - \alpha\gamma) = 0. \quad (24)$$

The structure of singularities of this equation depends on the values adopted by the roots $z_{1,2}$:

$$z_{1,2} = z_0 \pm \sqrt{\frac{\gamma + \delta z_0 + \varepsilon z_0^2}{\alpha + \varepsilon}}. \quad (25)$$

As compared with Eq. (3), it is seen that in the general case of distinct roots, $z_1 \neq z_2$, and if none of $z_{1,2}$ is zero, this equation has two additional regular singularities located at $z = z_1$ and $z = z_2$. If $z_1 = z_2$, we have only one extra singularity, located at $z = z_0$, which we suppose to be non-zero ($q \neq 0$). Besides, one of the extra singularities may coincide with the existing singularity $z = 0$ of the biconfluent Heun equation, however, we note that then the other additional singularity will necessarily be non-zero. Thus, if $q \neq 0$, Eq. (23) possesses at least one additional regular singularity located at a finite point of the complex $z$-plane, other than the origin.

Let this additional singularity be $z_1$, and consider a Frobenius solution of Eq. (23) in the neighborhood of this point:

$$w = \sum_{n=0}^{+\infty} a_n^{(z_1)} (z - z_1)^{\mu+n}. \quad (26)$$

Now, performing integration we arrive at the expansion

$$v(z) = C_1 + \sum_{n=0}^{+\infty} a_n^{(z_1)} \int z^{-1-\gamma}(z-z_1)^{\mu+n} dz = C_1 + \sum_{n=0}^{+\infty} a_n^{(z_1)} \frac{(-z_1)^{n+\mu}}{(z_1)^\gamma} B\left(-\gamma, 1+n+\mu; \frac{z}{z_1}\right), \quad (27)$$

which is indeed of the form of Eq. (10). Integrating once more, we arrive at the following final expansion of a solution of the biconfluent Heun equation in terms of combinations of the incomplete Beta functions:

$$u(z) = C_0 + C_1 z + \sum_{n=0}^{+\infty} a_n^{(z_1)} \frac{(-z_1)^{n+\mu}}{(z_1)^\gamma} \left( zB\left(-\gamma, 1+n+\mu; \frac{z}{z_1}\right) - z_1 B\left(1-\gamma, 1+n+\mu; \frac{z}{z_1}\right) \right). \quad (28)$$

We recall that here the integration constants $C_{0,1}$ are not arbitrary but should be appropriately chosen in order to complete the construction of the solution.

The coefficients $a_n^{(z_1)}$ of expansion (26) in general obey a five-term recurrence relation:

$$T_n a_n^{(z_1)} + S_{n-1} a_{n-1}^{(z_1)} + R_{n-2} a_{n-2}^{(z_1)} + Q_{n-3} a_{n-3}^{(z_1)} + P_{n-4} a_{n-4}^{(z_1)} = 0, \quad (29)$$

where

$$T_n = z_1(z_1 - z_2)(n+\mu)(n-1+\mu), \quad (30)$$

$$S_n = s_0 + s_1\mu_1 + s_2\mu_1^2, \quad R_n = r_0 + r_1\mu_1 + \mu_1^2, \quad Q_n = q_0 + q_1\mu_1, \quad \mu_1 = n + \mu, \tag{31}$$

$$P_n = \alpha + \varepsilon(n + \mu + 1 - \gamma), \tag{32}$$

the parameters $s_{0,1,2}$, $r_{0,1}$, $q_{0,1}$ being independent of $n$. Since the expansions (27),(28) are applicable only if $z_1 \neq 0$, this recurrence relation may involve four successive terms only if $z_1 = z_2$, i.e., if $\gamma + \delta z_0 + \varepsilon z_0^2 = 0$, or, equivalently, $\alpha^2\gamma + q\alpha\delta z_0 + q^2\varepsilon = 0$, see Eq. (25). In this case we have $z_{1,2} = z_0 = q/\alpha$, $T_n = 0$ and $S_n = z_0(n+\mu-1)(n+\mu-2)$, so that the series may left-hand side terminate if $\mu = 1$ or $\mu = 2$. Since these exponents differ by an integer, we may expect to have only one consistent power-series expansion corresponding to the greater exponent $\mu = 2$; the second solution may require a logarithmic term.

If $z_1 \neq z_2$ the series may left-hand side terminate if $\mu = 0$ or $\mu = 1$. In this case also it is generally expected to have only one consistent power-series expansion corresponding to the greater exponent $\mu = 1$.

Thus, for both cases, $z_1 = z_2$ and $z_1 \neq z_2$, the series may terminate from the right-hand side at some $N = 1, 2, \ldots$ if $P_N = 0$, that is if

$$\alpha = -\varepsilon(N + \mu + 1 - \gamma), \tag{33}$$

and additionally if $a_{N+1} = a_{N+2} = a_{N+3} = 0$ (in the case of a five-term relation) or $a_{N+1} = a_{N+2} = 0$ (in the case of a four-term relation).

## 3. Expansions in terms of the incomplete Gamma functions

Thus, starting from two equations obeyed by functions involving the first and the second derivatives of the biconfluent Heun function we have constructed two expansions of the solutions of the biconfluent Heun equation in terms of incomplete Beta functions. One of the expansions involves single Beta functions as expansion functions, while for the second series the expansion functions involve combinations of Beta functions with a linear in $z$ coefficient. The coefficients of the expansions obey four- and five-term recurrence relations, respectively. The constructed series apply to arbitrary set of the involved parameters of the biconfluent Heun equation except $\alpha = 0$, $q = 0$.

The same approach as above can be applied to construct series solutions in terms of other special functions. For instance, it is straightforwardly seen that applying the following differential equation obeyed by the function $v(z) = e^{\delta z} z^\gamma du/dz$:

$$v_{zz} + \left(\frac{1-\gamma}{z} - \delta + \varepsilon z - \frac{\alpha}{\alpha z - q}\right)v_z + \frac{P_3(z)}{z(\alpha z - q)}v = 0, \tag{34}$$

where $P_3$ is the cubic polynomial

$$P_3(z) = -\alpha\delta\varepsilon z^3 + (\alpha^2 + q\delta\varepsilon + \alpha\varepsilon(1-\gamma))z^2 - q(2\alpha + (2-\gamma)\varepsilon)z + q^2, \tag{35}$$

and applying a Frobenius solution for $v(z)$ in the neighborhood of the singular point $z = 0$:

$$v(z) = \sum_{n=0}^{\infty} c_n^{(0)} z^{\mu+n} \tag{36}$$

one can construct an expansion in terms of incomplete Gamma functions:

$$u(z) = C_0 + \int e^{-\delta z} z^{-\gamma} \left(\sum_{n=0}^{\infty} c_n^{(0)} z^{\mu+n}\right) dz = C_0 - \sum_{n=0}^{+\infty} \frac{c_n^{(0)}}{\delta^{n+\mu+1-\gamma}} \Gamma(n+\mu+1-\gamma; \delta z). \tag{37}$$

Another expansion in terms of incomplete Gamma functions is readily constructed if the differential equation obeyed by the function $v(z) = e^{\varepsilon z^2/2} z^\gamma du/dz$ is considered:

$$v_{zz} + \left(\frac{1-\gamma}{z} + \delta - \varepsilon z - \frac{\alpha}{\alpha z - q}\right)v_z + \frac{P_3(z)}{z(\alpha z - q)}v = 0, \tag{38}$$

where $P_3$ is now the following cubic polynomial

$$P_3(z) = -\alpha\delta\varepsilon z^3 + (\alpha^2 + q\delta\varepsilon)z^2 - \alpha(2q + \gamma\delta)z + q^2 + q\delta(\gamma - 1), \tag{39}$$

Then, again applying a Frobenius solution for $v(z)$ in the neighborhood of the singular point $z = 0$, we arrive at the expansion

$$u(z) = C_0 + \int e^{-\varepsilon z^2/2} z^{-\gamma} \left(\sum_{n=0}^{\infty} c_n^{(0)} z^{\mu+n}\right) dz = C_0 - \sum_{n=0}^{+\infty} \frac{c_n^{(0)}/2}{(\sqrt{\varepsilon/2})^{n+\mu+1-\gamma}} \Gamma\left(\frac{n+\mu+1-\gamma}{2}; \frac{\varepsilon z^2}{2}\right). \tag{40}$$

We conclude by noting another point supporting the observation concerning the usefulness of the application of the equations obeyed by the derivatives of the Heun functions. It is immediately seen from Eq. (22) that if its last term identically vanishes, then the solution of the problem is written in quadratures. We then arrive at the following solution to the initial biconfluent Heun equation (3):

$$u = C_1 e^{-\delta z - \varepsilon z^2/2} z^{-\gamma} + \frac{\gamma + \delta z + \varepsilon z^2}{\alpha z - q}\left(C_2 - C_1 \int e^{-\delta z - \varepsilon z^2/2} z^{-1-\gamma}(\alpha z - q) dz\right), \tag{41}$$

which is valid if $\alpha + \varepsilon = 0$ and $q^2 - \delta q - \alpha\gamma = 0$. Though this solution can be somehow guessed from Eq. (3), however, this derivation is clear, intuitive and straightforward.

## 4. Conclusions

Thus, we have demonstrated that the equations obeyed by different functions involving the derivatives of the Heun functions can be applied to generate expansions of the solutions of the Heun equations in terms of simpler special functions. We have shown this by constructing two expansions of the solutions of the biconfluent Heun equation in terms of the incomplete Beta functions and two expansions in terms of the incomplete Gamma functions. We note that under specific restrictions imposed on the parameters, the constructed series may terminate thus resulting in closed-form finite-sum solutions. Apart from this, the application of the equations obeyed by the derivatives of the Heun functions may be useful for construction of solutions written in quadratures. Though we have convinced in these conclusions discussing the particular case of the biconfluent Heun equation, these are general observations applicable to other Heun equations, as well as to other equations of more general type.

## Acknowledgments

This research has been conducted within the scope of the International Associated Laboratory (CNRS-France & SCS-Armenia) IRMAS. The work has been supported by the Armenian State Committee of Science (SCS Grant No. 15T-1C323). M. Gevorgyan thanks the Cooperation and Cultural Action Department (SCAC) of the French Embassy in Armenia for a doctoral grant.